\documentstyle[12pt]{article}
\begin{document}
\pagestyle{plain}
\title{The photo-electric effect in the bi-layer graphite}
\author{Miroslav Pardy\\
Department of Physical Electronics \\
Masaryk University \\
Kotl\'{a}\v{r}sk\'{a} 2, 611 37 Brno, Czech Republic\\
e-mail:pamir@physics.muni.cz}
\date{\today}
\maketitle
\vspace{50mm}

\begin{abstract}
We derive the differential probability of the photoelectric effect
realized at the very low temperature of double graphene  in the very strong 
magnetic field. The relation of this effect to the elementary particle
physics, nuclear physics and  Einstein gravity is mentioned. Our approach is the analogue of the Landau discovery of the diamagnetism,
where Landau supposed the parabolic dispersion relations for the model of diamagnetism.

\end{abstract}
\vspace{1cm}

\section{Introduction}

The photoelectric effect is a quantum electronic phenomenon in 
which electrons are emitted from matter after the absorption 
of energy from electromagnetic radiation.
Frequency of radiation must be above a threshold frequency, 
which is specific to the type of surface and material. 
No electrons are emitted
for radiation with a frequency below that of the threshold. 
These emitted electrons are also known as photoelectrons in this 
context. The photoelectric effect was theoretically explained by
Einstein who introduced the light quanta.
Einstein writes [1]: {\it In accordance with the
assumption to be considered here, the energy of light ray spreading
out from  point source is not continuously distributed over an
increasing space but consists of a finite number of energy quanta
which are localized at points in space, which move without dividing,
and which can only be produced and absorbed as complete units}.

The effect is also termed as the Hertz effect due 
to its discovery by Heinrich Rudolf Hertz in 1887. 

It is known  some prehistory of the photoelectric
effect beginning by  1839 when Alexandre Edmond Becquerel observed 
the photoelectric effect via an current when an electrode was  
exposed to light. Later in 1873, Willoughby Smith found that selenium 
is photoconductive.

The linear dependence on the
frequency was experimentally determined in  1915 when 
Robert Andrews Millikan showed that Einstein formula 

$$ \hbar\omega = \frac{mv^{2}}{2} + A \eqno(1)$$
was correct. Here $\hbar\omega$ is the energy of the impinging photon
and $A$ is work function of concrete material. The work function for
Aluminium is 4.3 eV, for Beryllium 5.0 eV, for Lead 4.3 eV, for Iron
4.5 eV, and so on [2]. The  work function concerns the
surface photoelectric effect where the photon is absorbed by an
electron in a band. The theoretical determination of the work function 
is the problem of the solid state physics. On the other hand, there is
the so called atomic photoeffect [3], where the ionization
energy plays the role of the work function. The system of the
ionization energies is involved in the tables of the solid
state physics. The work fuction of graphene, or, work fuction  of the Wigner crystal in graphene was never 
determined, and it is the one of the prestige problem of the contemporary experimental and theoretical
graphene physics and the Wigner crystal physics. 

The formula (1) is the law of conservation of energy.
The classical analogue of the equation (1) is the motion of 
the Robins ballistic pendulum in the resistive medium.

The Einstein ballistic principle is not valid inside of the
blackbody. The Brownian motion of electrons in this cavity is caused
by the repeating Compton process $\gamma + e\to \gamma + e$ and not by
the ballistic collisions. The diffusion constant for electrons must be
calculated from the Compton process and not from the Ballistic
process. The same is valid for electrons immersed into the cosmic relic
photon sea. 
 
The idea of the existence of the Compton
effect is also involved in the Einstein article. He writes [1]: 
{\it The possibility should not be excluded, however, that electrons 
might receive their energy only in part from the light
quantum}. However, Einstein was not sure, a priori, that his idea of
such process is realistic. Only Compton proved the reality of the
Einstein statement.

Eq. (1) represents so called one-photon photoelectric effect, which is
valid for very weak electromagnetic waves.  At
present time of the laser physics, where the strong electromagnetic
intensity is possible,  we know that so called multiphoton
photoelectric effect is possible. Then, instead of equation (1) we can
write

$$ \hbar\omega_{1} +  \hbar\omega_{2} + ...  \hbar\omega_{n} = 
\frac{mv^{2}}{2} + A .\eqno(2)$$

The time lag between the incidence
of radiation and the emission of a photoelectron is very small, 
less than $10^{-9}$ seconds. 

As na analogue of the equation (2), the multiphoton Compton effect is 
also possible:

$$\gamma_{1} + \gamma_{2} + ... \gamma_{n} + e  
\rightarrow \gamma + e ,\eqno(3)$$
and two-electron, three-electron,... n-electron  photoelectric effect is 
also possible [3]. To our knowledge the Compton process
with the entangled photons was still not discovered and elaborated. On
the other hand,  there is the deep inelastic Compton effect in the high
energy particle physics. 

Einstein in his paper [1] introduced the term  "light quanta"  called
"photons" by chemist G. N. Lewis, in 1926. Later Compton, in his famous
experiment proved that light quanta have particle properties, or, 
photons are  elementary particles.  

At present time the attention of physicists is concentrated to
the planar physics at zero temperature and in the strong magnetic field. 
Namely, in  the graphene physics which is probably new revolution 
in this century physics with assumption that graphene is the silicon of this century. 

In  2004, Andre Geim, Kostia Novoselov [4]
and co-workers at the University of Manchester in the UK  delicately 
cleaving a sample of graphite with sticky tape produced something
that was long considered impossible: a sheet of crystalline
carbon just one atom thick, known as graphene.
Geim's group was able to isolate graphene, and was
able to visualize the new crystal using a simple optical microscope.
Graphene is the benzene ring ($C_{6}H_{6}$) stripped out from their H-atoms. It is allotrope of carbon because carbon
can be in the crystalline form of graphite, diamant, fulleren ($C_{60}$) and carbon nanotube. 

Graphene unique properties arise from the collective
behaviour of electrons. The electrons in graphene are governed 
by the Dirac equation. The behavior of electrons in graphene
was first predicted in 1947 by the Canadian theorist
Philip Russell Wallace.  

The Dirac fermions in graphene carry one unit of
electric charge and so can be manipulated using electromagnetic
fields. Strong interactions
between the electrons and the honeycomb lattice of carbon atoms mean that the dispersion relation is linear and 
given by $E = vp$, $v$ is called the Fermi-Dirac velocity, $p$ is momentum of a quasielectron. The energy quantization of the electron in the bi-layer graphene in magnetic field is $E_{n} \sim \sqrt{n(n-1)}$ and the dispersion relation is quadratic. 
 The Dirac equation in graphene physics is used for so called quasispin generated by the honeycomb lattice.
The parabolic dispersion relation  valid for 2-layer graphite  means it is possible to use the  Schr\"odinger equation 
for the calculation of photoeffect in 2-layer graphite.

Our calculation of the photoelectric effect is applicable not only for the 2-layer graphite but also for the Wigner crystals. 
A Wigner crystal is the crystalline  phase of electrons first predicted by Eugene Wigner in 1934 [5], who was probably motivated  by the electron quantum states forming the Landau diamagnetism. In other words this is a gas 
of electrons moving in 2-dimensional  or 3-dimensional  neutralizing background as a lattice if the electron density is less than a critical value 
potential energy $E_{p}$ dominating  the kinetic energy $E_{k}$ at low densities. Or, $E_{p} > E_{k}$.
To minimize the potential energy, the electrons form a triangular lattice in 2D and body-centered cubic lattice in 3D.
A crystalline state of the 2D electron gas can also be realized by applying a sufficiently strong magnetic field. And this is a case of monolayer graphite (graphene) in the magnetic field. It is evident that the  2-layer graphite involves the Wigner crystal too [6]. 

We derive the differential probability of the photoelectric effect
realized at the very low temperature graphene in the very strong 
magnetic field. The relation of this effect to the elementary particle
physics of LHC, nuclear physics and Einstein gravity is mentioned.

\section{The quantum theory of the photoelectric effect in the 2D 
films at zero temperature and strong magnetic field}

Electrons in graphene respect the linear spectrum and they are described by the Dirac equation. The electron dispersion relations in the 2-layer graphite was proved to be parabolic and we can calculate the photoelectric effect using the Schr\"odinger equation. In general words the photoelectric effect on graphene can give us information on the state of electrons in general in N-layer graphite.

The photoelectric effect in graphene is presented here 100 years after well know Einstein article [1]. While the experimental investigation of the photoelectrical effect was performed in past many times, the photoelectric effect in graphene is still the missing experiment in the graphene physics, the electronics  of the future. Nevertheless the photoeffect is the brilliant method for investigating the graphene and the Wigner crystals.     

The quantum mechanical description of the photoeffect is realized as 
the nonrelativistic, or relativistic and it is described in many textbooks on
quantum mechanics. Let us apply the known nonrelativistic elementary quantum theory of
the photoeffect to the 2D structures with the 2-layer graphite.

The main idea of the quantum mechanical description of the photoeffect
is that it must be described by the appropriate S-matrix element
involving the interaction of atom with the impinging photon with the
simultaneous generation of the electron, the motion of which can be
described approximately by the plane wave

$$\psi_{\bf q} = \frac{1}{\sqrt{V}}e^{i{\bf q}\cdot{\bf x}},\quad  
{\bf q} = \frac{\bf  p}{\hbar},  \eqno(4)$$ 
where ${\bf p}$ is the momentum of the ejected electron. 
We suppose that magnetic field is applied locally to the carbon film, so, in a sufficient distance from it the 
wave function is of the form of the plane wave (4). This situation has an analog in the classical atomic effect discussed in monograph [7]. However, if the photon energy only just exceeds the ionization energy $I$ of atom, then we cannot used the plane wave approximation but the wave function of the continuous spectrum. 

The probability of the emission of electron by the electromagnetic
wave is of the well-known form [7]:

$$dP = \frac{e^{2}p}{8\pi^{2}\varepsilon_{0}\hbar m\omega}
\left|\int e^{i({\bf k} - {\bf q})\cdot{\bf x}}({\bf e}\cdot\nabla)
\psi_{0}dxdydz\right|^{2}d\Omega  = C|J|^{2}d\Omega, \eqno(5)$$
where the interaction  for absorption of the
electromagnetic wave is normalized  to {\it one photon in the unit volume},  
${\bf e}$ is the polarization of the impinging photon,  
$\varepsilon_{0}$ is the dielectric constant of vacuum, $\psi_{0}$ is the basic state of and atom. We have denoted
the integral in $||$ by $J$ and the constant before $||$ by C.  

In case of the 2D low temperature system in the strong magnetic field
the basic function is so called Laughlin function [8],
which is of the very sophisticated Jastrow form.
We consider the case with electrons in magnetic field  as an analog of the Landau diamagnetism. 
So, we take the basic function $\psi_{0}$ for one
electron in the lowest Landau level, as 

$$\psi_{0} = \left(\frac{m\omega_{c}}{2\pi\hbar}\right)^{1/2}
\exp\left(-\frac{m\omega_{c}}{4\hbar}(x^{2} + y^{2})\right), \eqno(6)$$ 
which is solution of the Schr\"odinger equation in the magnetic field
with potentials ${\bf A} = (-Hy/2, -Hx/2, 0)$,  $A_{0} = 0$
[9]:

$$\left[\frac{p_{x}^{2}}{2m} + \frac{p_{y}^{2}}{2m} - \frac{m}{2}
\left(\frac{\omega_{c}}{2}\right)^{2}(x^{2} + y^{2})\right]\psi = E\psi.
\eqno(7)$$ 

We have supposed that the motion in the z-direction is zero and it
means that the wave function $\exp[(i/\hbar) p_{z}z] = 1$. 

So, The main problem is to calculate the integral 

$$J = \int e^{i({\bf K}\cdot{\bf x})}({\bf e}\cdot\nabla)\psi_{0}dxdydz;\quad
{\bf K} = {\bf k} - {\bf q}. \eqno(8)$$
with the basic Landau function $\psi_{0}$ given by the equation (6).

Operator $(\hbar/i)\nabla$ is Hermitean and it means we can rewrite 
the last integrals as follows:

$$J = \frac{i}{\hbar}{\bf e}\cdot\int
 \left[\left(\frac{\hbar}{i}\nabla\right)
 e^{i({\bf K}\cdot{\bf x})}\right]^{*}\psi_{0}dxdydz, \eqno(9)$$
which gives 

$$J = i{\bf e}\cdot{\bf K}\int  e^{-i({\bf K}\cdot{\bf x})}\psi_{0}dxdydz, 
\eqno(10)$$

The integral  in eq. (10) can be transformed using the cylindrical 
coordinates with 

$$ dxdydz = \varrho d\varrho d\varphi dz, \quad \varrho^{2} = x^{2} + y^{2}\eqno(11)$$  
which gives  for vector 
${\bf K}$ fixed on the axis z with ${\bf K}\cdot{\bf x} = Kz$ and
with physical condition  ${\bf e}\cdot{\bf k} = 0$, expressing the
physical situation where polarization is perpendicular to the
direction of the wave propagation. So, 

 $$J = (i) ({\bf e}\cdot{\bf q})\int_{0}^{\infty}\varrho d\varrho
\int_{-\infty}^{\infty}dz\int_{0}^{2\pi}d\varphi e^{-iKz}\psi_{0}.\eqno(12)$$

Using 
$$\psi_{0} = A\exp\left(-B\varrho^{2}\right);\quad
A = \left(\frac{m\omega_{c}}{2\pi\hbar}\right)^{1/2}; \quad  
B = \frac{m\omega_{c}}{4\hbar}.  \eqno(13)$$
 The integral (12) is then  

 $$J = (-\pi i)\frac{A}{B} ({\bf e}\cdot{\bf q})
\int_{-\infty}^{\infty}e^{-iKz}dz
= (-\pi i)\frac{A}{B} ({\bf e}\cdot{\bf q})(2\pi)\delta(K).\eqno(14) $$
Then,

$$dP = C|J|^{2}d\Omega = 4\pi^{4}\frac{A^{2}}{B^{2}}C 
({\bf e}\cdot{\bf   q})^{2}\delta^{2}(K)d\Omega. \eqno(15)$$

Now, let be the angle $\Theta$  between direction ${\bf k}$
and direction ${\bf q}$, and let be the angle $\Phi$ between
planes $({\bf k},{\bf q})$ and $({\bf e},{\bf k})$. Then, 

$$({\bf e}\cdot{\bf q})^{2} = q^{2}\sin^{2}\Theta\cos^{2}\Phi.\eqno(16)$$

So, the differential probability of the emission of photons from the
2-layer graphite (double graphene) in the strong magnetic field is as follows:

$$dP = \frac{4e^{2}p}{\pi \varepsilon_{0} m^{2}\omega \omega_{c}}\left[q^{2}
\cos^{2}\Theta\sin^{2}\Phi\right]\delta^{2}(K)d\Omega; \quad \omega_{c} = \frac{|e|H}{mc}. \eqno(17)$$

We can see that our result differs form the result for the
original photoelectric effect [7] which involves still the term

$$\frac{1}{(1 - \frac{v}{c}\cos\Theta)^{4}}, \eqno(18)$$
which means that the most intensity of the classical photoeffect
is in the direction of the electric vector of the electromagnetic wave 
$(\Phi = \pi/2, \Theta = 0)$. While the nonrelativistic solution of
the photoeffect in case of the Coulomb potential was performed by Stobbe
[10] and  the relativistic calculation by Sauter [11], the general 
magnetic photoeffect (with electrons moving in the magnetic field and forming atom) was not still performed in a such simple form.
The delta term $\delta\cdot\delta$ represents the conservation law
$|{\bf k} - {\bf q}| = 0$ in our approximation. While the product
$\delta\cdot\delta$ forms no problem in the quantum field theory, its
mathematical  meaning is not well defined in the so called theory of
the generalized functions.    
 
\section{Discussion}

The article is in a some sense the preamble to the any conferences of ideas related to the photoeffect on graphene, bi-layer graphite, n-layer graphite and on the Wigner crystals which are spontaneously formed in graphite structures, or,  in other structures.
We have considered the photoeffect in the planar crystal at zero
temperature and in the very strong magnetic field. 
We calculated only the process which can be approximated by the Schr\"odinger equation for an electron orbiting in magnetic field. 

At present time, the most attention in graphene physics is devoted to the conductivity of a graphene with the goal to invent new MOSFETs and new transistors for new computers. However, we do not know, a priory, how many discoveries are involved in the 
investigation of he photo-electric effects in graphene.

We did not consider the relativistic description in the constant
magnetic field. Such description can be realized using the so called
Volkov solution of the Dirac equation in the magnetic field instead of
the plane wave . The explicit form of such solution was used by
Ritus [12], Nikishov [13] and others, and by author [14], [15], [16],
in the different situation. For instance, for the
description of the electron in the laser field, synchrotron radiation, 
or, in case of the massive photons leading to the Riccati equation [15]. 

The Volkov solution for an electron moving in the constant magnetic field 
is the solution of the Dirac equation with the following four potential

$$A_{\mu} = a_{\mu}\varphi; \quad \varphi = kx; \quad k^{2} = 0. \eqno(19)$$

It follows from equation (19) that $F_{\mu\nu} = \partial_{\mu}A_{\nu} -
\partial_{\nu}A_{\mu}  = - a_{\mu}k_{\nu} + a_{\nu}k_{\mu} = const.$, which
means that electron moves in the constant electromagnetic field with the
components ${\bf E}$ and ${\bf H}$. The parameters $a$ and $k$ can be chosen
in a such a way that ${\bf E} = 0$. So, the motion of electron is performed in
the constant magnetic field.

The  Volkov solution [17] of
the Dirac equation for an electron moving in a field of a plane wave
was derived in the form [14],[15],[18]: 

$$\psi_{p} = \frac {u(p)}{\sqrt{2p_{0}}}
\left[1 + e \frac {(\gamma k)
(\gamma A(\varphi))}{2kp}
\right] \exp \left[(i/\hbar)S \right]\eqno(20)$$
and $S$ is an classical action of an electron moving in the 
potential $A(\varphi)$ [19].

$$S = -px - \int_{0}^{kx}\frac {e}{(kp)}\left[(pA) - \frac {e}{2}
(A)^{2}\right]d\varphi. \eqno(21)$$

It was shown that for the potential (19) the Volkov wave function 
is [19]:

$$\psi_{p} = \frac {u(p)}{\sqrt{2p_{0}}}
\left[1 + e \frac {(\gamma k)
(\gamma a)}{2kp}\varphi
\right] \exp \left[(i/\hbar)S \right]\eqno(22)$$
with

$$S = -e\frac {ap}{2kp}\varphi^{2} + e^{2}\frac {a^{2}}{6kp}\varphi^{3} -
px. \eqno(23)$$

The basic function is also relativistic and it was derived in the form 
[20]. 

$$\Psi({\bf x},t) = \frac{1}{L}\exp\{-\frac{i}{\hbar}\epsilon Et +
ik_{2}y + ik_{3}z\}\psi; \quad
\psi = \left(\begin{array}{c}C_{1}u_{n-1}(\eta)\\
iC_{2}u_{n}(\eta)\\
C_{3}u_{n-1}(\eta)\\
iC_{4}u_{n}(\eta)
\end{array}\right), 
\eqno(24)$$
where $\epsilon = \pm 1$ and the spinor components $u$ and the 
coefficients $C_{i}$ are defined in the Sokolov et al. monograph [20].

$$u_{n}(\eta) = \sqrt{\frac{\sqrt{2\gamma}}{2^{n}n!}\sqrt{\pi}}\; 
e^{-\eta^{2}/2}H_{n}(\eta)\eqno(25)$$
with 

$$H_{n}(\eta) = (-1)^{n}e^{\eta^{2}}\left(\frac{d}{d\eta}\right)^{n}
e^{-\eta^{2}},\eqno(26)$$

$$\eta = \sqrt{2\gamma}\;x + k_{2}/\sqrt{2\gamma}; \quad 
\gamma = eH/2c\hbar .\eqno(27)$$

The coefficients $C_{i}$ are defined in the Sokolov et al. 
monograph [20]. So, our elementary approach can be generalized. 

We have here considered the situation which is an analog of the 
ionization process in atom. Namely:

$$\hbar\omega + (e\; in\; atom) \rightarrow  (atom\; minus\; e) +  e. 
\eqno(28)$$

In other words,  we have calculated the process with the following equation:

$$\hbar\omega + (e\; in \;G)  \rightarrow (G\; minus\; e) + e,\eqno(29)$$ 
where $G$ is the double graphene and the electron in the crystal  is considered 
as the elementary particle with all atributes of electron as an
elementary particle and not the quasielectron which can be also 
considered in the solid state physics. Quasielectron is the product of
the crystalline medium and as a such it can move only in the
medium. The existence of the quasielectron in vacuum is not
possible. However, the process where the photon interacts with the
quasielectron and after some time the quasielectron decays in
such a way that the integral part of the 
decay product is also electron, is possible and till
present time the theory of such process was not elaborated.

The new experiments are
necessary in order to verify the photoelectric equation in graphene. 
At the same time it will be the dilemma if  the 
process is ballistic,  or not. The ballistic interaction of photon
with electron is not possible in vacuum for point-like electrons. The
experiments in CERN, Hamburg, Orsay, (The L3 collaboration, CELLO
collaboration, ALEPH collaboration, ...) and other laboratories 
confirmed that there are no excited states of electron in
vacuum. In other words, the ballistic process in vacuum with electrons 
was not confirmed. It means that electron in the standard model of
elementary particles is a point particle. It seems
that the ballistic process is possible only in a medium. 
We know, that electron accelerated in PASER [21] by 
the ballistic method is accelerated 
for instance in $\rm CO_{2}$. On the other hand, the
photo-desintegration of nuclei involves the ballistic process
[22]. The interaction of light with carbon $C_{60}, C_{70}, C_{80} ... $
involves also the ballistic process with photons [23, 24].
   
The photoelectric effect at zero temperature can be realized only by very
short laser pulses, because in case of the continual laser irradiation 
the zero temperature state is not stable. Only very short
pulses can conserve the zero temperature 2D system. 
 
The interaction of light with quasielectron is substantially new process 
which will be obviously one of the future problem of the norelativistic and relativistic  
graphene physics, or, the graphene quantum optics. Such interaction is the analogue of the interaction of light with phonons (quanta of the elastic waves), magnons (quanta of the magnetic waves), plasmons (quanta of the plasmatic waves), polarons (quanta of the bound states of electrons and their accompanying polarization waves), polaritons (quanta of the polarization waves), dislocons (quanta of the dislocation waves), excitons (quanta of the excitation waves), Cooper pairs (two-electron system  of electrons with the opposite spins in superconductor), and so on.

The graphene can be deformed in such a way that the metric of the deformed sheet is the Riemann one. However, the Riemann metric of general relativity is the metric of the deformed 4D sheet as was proposed by Sacharov [25]. So, in other words,  there is the analogy between deformation and Einstein gravity. The discussion on this approach was presented also by author [26]. The deformed graphene  obviously leads to the modification of the photoeffect in graphene and it can be used as the introduction to the theory of the photoelectric effect influenced by the gravitational field.   

The information on the photelectrical effect in graphene and also the
elementary particle interaction with graphene is necessary not only in
the solid state physics, but also in the elementary particle physics in
the big laboratories where graphene can form the substantial components of
the particle detectors. The graphene can be probably used as the
appropriate components in the solar elements, the anode and cathode
surface in the electron microscope, or, as the medium of the memory hard
disks in the computers. While the last century economy growth was based on the Edison-Tesla electricity, the economy growth in this century will be 
obviously based on the graphene physics.  We hope that these perspective ideas will be considered at the universities and in the physical laboratories.

\vspace{7mm}

\noindent
{\bf References}

\vspace{7mm}

\noindent 
1. Einstein, A.:  Annalen der Physik {\bf 17}, 132 (1905); The English
translation is in: AJP {\bf 33}, No. 5 May  367 (1965) \\[2mm] 
2. Rohlf, J. W.:  Modern Physics from $\alpha$ to $Z^{0}$,
John Willey \& Sons, Inc., New York, (1994)\\[2mm]
3. Amusia, M. Ya.:  Atomic photoeffect, Nauka, Moscow, (1987) (in
Russian).\\[2mm]
4. Kane, C. L.: NATURE  {\bf 438}, November  168 (2005) \\[2mm]
5. Wigner, E. P.:  Phys. Rev. {\bf 46}, 1002 (1934) \\[2mm]
6. C$\rm \hat{o}t\acute{e}$ R.,  Jobidon J. -T. and  Fertig H. A, arXiv: cond-mat.mes-hall/0806.0573v1 \\[2mm]
7. Davydov, A. S.:  Quantum mechanics, Second Ed., Pergamon Press, Oxford, New York, ...  (1976)\\[2mm]
8. Laughlin, R. B.: Phys. Rev. Lett.  {\bf 50}, No. 10 1395 (1983) \\[2mm]
9. Drukarev, G. F.:   Quantum mechanics, St. Petersburgh University, (1988) (in Russsian)\\[2mm]
10. Stobbe, M.: Ann. der Phys. {\bf 7}, 661 (1930)\\[2mm]
11. Sauter, F.: Ann. der Phys.  {\bf 9}, 217 (1931)\\[2mm]
12. Ritus, V. I.: Trudy FIAN  {\bf 111}, 5 (1979) (in Russian) \\[2mm]
13. Nikishov, A. I.:Trudy FIAN  {\bf 111}, 152 (1979) (in Russian)\\[2mm]
14. Pardy, M.: IJTP {\bf 42},(1) 99 (2003)\\[2mm]
15. Pardy, M.:  IJTP {\bf 43},(1) 127 (2004)\\[2mm]
16. Pardy, M.: arXiv:hep-ph/0703102v1\\[2mm]
17. Volkov, D. M.:  Zeitschrift f\"ur Physik {\bf 94}, 250 (1935)\\[2mm]
18. Berestetzkii, V. B., Lifshitz, E. M. and Pitaevskii, L. P.: 
 Qantum electrodynamics, Moscow, Nauka, (1989) (in Russian) \\[2mm]
19. Landau, L. D. and Lifshitz E. M.:  The classical
theory of fields, 7-th ed., Moscow, Nauka, (1988) (in Russian) \\[2mm]
20. Sokolov, A. A. and   Ternov I. M.: Relativistic electron, 2nd ed., 
Moscow, Nauka, (1983) (in Russian)\\[2mm]
21. Banna, S.:  Berezovsky V. and  Sch\"achter L., Experimental evience
for particle acceleration by stimulated emission of radiation, AIP
Conference Proc. {\bf 877} 51 (2006) \\[2mm]
22. Levinger, J. S.: Nuclear photo-desintegration, Oxford
University Press, (1960). (in Russian)\\[2mm]
23. Scully, S. W. J.,  Emmons, E. D.,  Gharaibeh, M. F. and  Phaneuf, R. A.; 
Kilcoyne, A. L. D. and  Schlachter, A. S. ; 
Schippers, S.  and  M\"uller, A. ;  Chakraborty, H. S.,  Madjet, M. E. and  Rost, J. M.:
Phys. Rev. Lett. {\bf 94}, 065503 (2005) \\[2mm]
24. Physics News Update, Number {\bf 722}, Story \#1 (2005) \\[2mm]
25. Sacharov, A. D.: Dokl. Akad. Nauk SSSR {\bf 177},  70 (1967) \\[2mm]
26. Pardy, M.: in: Spacetime Physics Research Trends. 
Horizons in World Physics, Volume {\bf 248}, ISBN 1-59454-322-4 111 (2005)
\end{document}